\documentclass[twocolumn,aps,prl,superscriptaddress,showpacs]{revtex4}
\usepackage{hyperref}
\usepackage{graphicx}
\usepackage{amsmath}
\usepackage{epsfig}
\usepackage[utf8]{inputenc}

\newcommand{\bra}[1]{\mathop{\left\langle #1 \right|}\nolimits}
\newcommand{\ket}[1]{\mathop{\left| #1 \right\rangle}\nolimits}

\begin{document}
\title{A robust quantum receiver for phase shift keyed signals}

\author{Christian~R.~M\"{u}ller}
\affiliation{Max Planck Institute for the Science of Light, Erlangen, Germany}
\affiliation{Friedrich-Alexander-University Erlangen-N\"{u}rnberg (FAU), Department of Physics, Germany}

\author{Christoph~Marquardt}
\affiliation{Max Planck Institute for the Science of Light, Erlangen, Germany}
\affiliation{Friedrich-Alexander-University Erlangen-N\"{u}rnberg (FAU), Department of Physics, Germany}

\date{\today}

\begin{abstract}
The impossibility of perfectly discriminating non orthogonal quantum states imposes far-reaching consequences both on quantum and classical communication schemes. 
We propose and numerically analyze an optimized quantum receiver for the discrimination of phase encoded signals. 
Our scheme outperforms the standard quantum limit and approaches the Helstrom bound for any signal power. 
The discrimination is performed via an optimized, feedback-mediated displacement prior to a photon counting detector. 
We provide a detailed analysis of the influence of excess noise and technical imperfections on the average error probability. 
The results demonstrate the receiver's robustness and show that it can outperform any classical receiver over a wide range of realistic parameters. 
\end{abstract}

\pacs{03.67.Hk, 03.65.Ta, 42.50.Ex}

\maketitle
The exponentially increasing global data traffic necessitates the optimization of all telecommunication technology. 
Along with these developments individual components like receivers are gradually approaching physical limits such that quantum mechanical approaches are gaining significance. 
In optical communications, the information to be exchanged among the communicating parties is encoded into one or more parameters of the light field.
Besides conventional encoding protocols like intensity modulation, also coherent encodings have found their way into commercial application in the course of the last few year. 
The prepared signal states are transmitted through an optical channel and detected by a receiver which retrieves the encoded information via an adequate measurement. 
Coherent states are of outstanding importance for optical communications, since they are loss-tolerant and readily produced signal carriers. 
The ultimate channel capacity for a lossy channel can already be achieved using merely coherent states combined with appropriate classical coding and quantum optimal measurements \cite{Giovannetti04}. 
A fundamental class of coherent signal codings is phase shift keying (PSK). The quadrature phase shift keying (QPSK) alphabet, for instance, comprises four coherent states $\{\ket{\alpha}, \ket{i \alpha}, \ket{-\alpha},\ket{-i \alpha}\}$ of identical amplitude $|\alpha|$, equally separated by a phase-shift of $\pi/2$ (see Fig.\ref{BondurantOD}). 
Ideally, this alphabet allows for the transmission of two bits of information per signal state. 
However, coherent states are mutually non-orthogonal and hence cannot be discriminated perfectly. 
The underlying quantum uncertainties ultimately limit the channel capacity in optical communication \cite{Giovannetti04}, but they also allow for applications such as quantum key distribution (QKD) \cite{Bennett92, Scarani09, Weedbrook12, Lorenz04, Sych10, Leverrier09}. 
The minimal error probability for the discrimination of non-orthogonal states is usually referred to as the Helstrom bound \cite{Helstrom67, Helstrom76, Bergou04}.

Fundamental studies of physical receiver implementations reaching the Helstrom bound were first considered by Dolinar \cite{Dolinar73} and Kennedy \cite{Kennedy73}. 
More recently, also optimal detection operators and strategies were investigated \cite{HirotaBan96, Nair12, Banaszek99}. In conventional receivers, i.e. heterodyne or dual-homodyne detection, the error probability is lower bounded by shot noise, which constitutes the standard quantum limit (SQL). 
A QPSK receiver surpassing the SQL for mean photon numbers $n > 0.7$ and based on adaptive displacements, direct detection and feedback has been proposed by Bondurant \cite{Bondurant93}. With the advent of fast feedback electronics, variants of this scheme could be demonstrated experimentally \cite{Becerra11, Becerra13NatPhot}. 
Alternative approaches carry out the the discrimination by multiplexing the states into several branches, applying (nulling) displacements and detecting the partial states via on-off detection or photon number resolving detection \cite{Izumi12, Becerra13NatComm, Assalini11, Sych13, Nair14}. Squeezing assisted state discrimination \cite{Izumi13} was also considered and shown to allow for error probabilities outperforming the SQL.

The aforementioned QPSK receivers approach the Helstrom bound for large signal amplitudes. However, their performance in the regime of small mean photon numbers, i.e. $n \lesssim 1$ is severely limited such that they cannot achieve error probabilities below the SQL in this regime.
An approach that allows to surpass the SQL for any signal power is to use a hybrid receiver \cite{Mueller12} that discriminates the states in two steps via a homodyne detector and a displacement receiver. However, this hybrid receiver only provides a moderate advantage over the conventional heterodyne detector.

In this article we propose an adaptive, feedback-mediated displacement receiver for the discrimination of phase shift keyed coherent states. 
We show that our scheme surpasses the standard quantum limit significantly for any signal power and offers an unprecedentedly small error probability in the regime of weak signals. 
Our receiver minimizes the signals' average error probability by applying an optimized displacement instead of the conventional displacement that shifts one of the signals exactly to the vacuum state.
In schemes relying on such a perfect nulling, only the error probability of the nulled signal is minimized, the sum of the individual error probabilities, however, is not minimized. 
The resulting imbalance between the individual error probabilities in these schemes translates into a biased final hypothesis which is most pronounced in the regime of weak signal powers. By reducing this bias our optimized displacement receiver allows for a significantly smaller error probability. 
We discuss different realizations of the optimized displacement scheme and investigate their robustness against excess noise and technical imperfections of the photon detector. 
It should be noted here, that an alternative approach to optimize the average error probability is to adjust the \textit{a priori} probabilities of the signal states as was recently proposed by the Sasaki group \cite{Sasaki13}. 

The article is structured as follows. 
In section I we provide a basic introduction to the receiver's state discrimination strategy, which establishes the foundation for the more detailed discussions in the subsequent sections. 
In section II we discuss the adaptive optimized displacement receiver and compare the performance for two different signal probing strategies - a cyclic probing strategy and a more advanced strategy based on Bayesian inference of the individual \textit{a posteriori} probabilities.  
Finally, in section III we discuss the robustness of the scheme against excess noise, detector dead time, dark counts and finite quantum efficiency of the detectors.

\section{I. Adaptive quantum receivers}
In coherent optical communications, the signal states are defined as pulses of coherent light with a certain temporal extent $T$. To facilitate the theoretical description of the receiver scheme, we will henceforth assume the signal states to be prepared with a rectangular pulse shape and with a normalized temporal extent $T=1$. This assumption does not limit the generality of the proposed scheme, as any other pulse shape can straightforwardly be taken into account by an additional weighting factor.  
The measurement of a signal state is only complete after the entire pulse has entered the detector. 
During the measurement time, information on the signal is continuously acquired. 
For the task of quantum state discrimination, this offers the possibility for an on-the-fly adaption of free parameters of the receiver based on the available partial information. 

An important representative of this adaptive receiver class is the Bondurant receiver \cite{Bondurant93}. 
It is a receiver for the discrimination of QPSK signals, $\ket{\alpha_k}=\ket{\alpha\,\mathrm{e}^{i k \pi/2}},\,k \in \{0,1,2,3\}$, composed of an adaptive displacement and an on-off detector. 
The initial displacement is chosen to null the first signal state, i.e. the state $\ket{\alpha_1}$ is displaced to the vacuum state $\ket{0}$. 
As the vacuum state is an eigenstate of the photon number basis with eigenvalue $0$, any photon detection unambiguously excludes the state $\ket{\alpha_1}$ from the set of compatible hypotheses. 
Consequently, the displacement is changed to probe for the subsequent state $\ket{\alpha_2}$ and so forth. 
The final hypothesis is determined by the state that is displaced to the vacuum state after the hindmost detection event. 
The error probability is solely depending on the signal's mean photon number and for bright signals $|\alpha|^2\gg 1$ the Bondurant strategy provides the same exponential scaling as the Helstrom bound. 
In the regime of weak signals, however, the performance of the Bondurant receiver deteriorates rapidly. 
At signal powers $|\alpha|^2 \lesssim 0.7$ the error probability exceeds the standard quantum limit, i.e. the quantum receiver has no advantage over classical schemes. This shortcoming is mainly a consequence of the non-linear relation between the signal's mean photon number $n$ and the probability for the detection of at least one photon $p_{click} = 1- \exp(-n)$.
For small signal powers the probability to register even just a single photon detection tends to zero, which results in a disadvantageous bias of the final hypothesis towards the initially probed signal state $\ket{\alpha_1}$.

Deeper insight into the detrimental nature of this bias can be gained by analyzing the particular case of the binary phase shift keying alphabet (BPSK) $\ket{\alpha_k}= \ket{\alpha\,\mathrm{e}^{i k \pi}}, k \in \{0,1\}$.
A single photon detector based receiver for the BPSK alphabet is the Kennedy receiver \cite{Kennedy73}. It displaces one of the two signal states 
to the vacuum state $\ket{0}$, while the other state is consequently displaced to $\ket{2\alpha}$. 
Subsequently, the displacement is kept constant and the signal is measured by an on-off detector. 
The final hypothesis for the input state is drawn from the observation whether or not at least one photon has been detected. 
The state displaced to the vacuum will never generate an erroneous detection event. 
Consequently, any errors in this configuration stem from measurements in which the bright state $\ket{2\alpha}$ was projected to the vacuum state and hence failed to excite a detection event.   
The final hypothesis is therefore biased towards the nulled state. 

Obviously, the error probability of the nulled signal is optimized in the Kennedy receiver. 
However, the sum of the individual error probabilities is not necessarily minimized. 
In the regime of weak signals the error probability can be reduced significantly by merely increasing the displacement amplitude to an optimal value. 
At this configuration the additional errors stemming from the dim signal not being in the vacuum state are overcompensated by the increased photon detection probability of the brighter state \cite{Takeoka08, Wittmann08}.

The situation is considerably similar in the case of higher-dimensional PSK alphabets like the Bondurant receiver. With decreasing signal amplitude, the probability for any of the not nulled signal states to generate even just a single detection event tends to zero and the final hypothesis is strongly biased towards the initially nulled state $\ket{\alpha_1}$.  
Just as for the binary alphabet, an optimized displacement relaxes the bias and thus allows for significantly smaller error probabilities as will be described in the following section.

\section{II. Implementation of the optimized displacement receiver}
In this section the state discrimination protocol of the optimized displacement receiver is described in detail. 
In order to keep the theoretical description universal, the dimension of the phase shift keyed alphabet $M$ is not fixed in the formal expressions but remains a free parameter of the protocol. 
However, for the purpose of illustrating the discrimination procedure and for the sake of providing demonstrative curves of the resulting error probabilities, we evaluate the practical example of the QPSK alphabet.

\begin{figure}[htb]
\includegraphics[width=8cm]{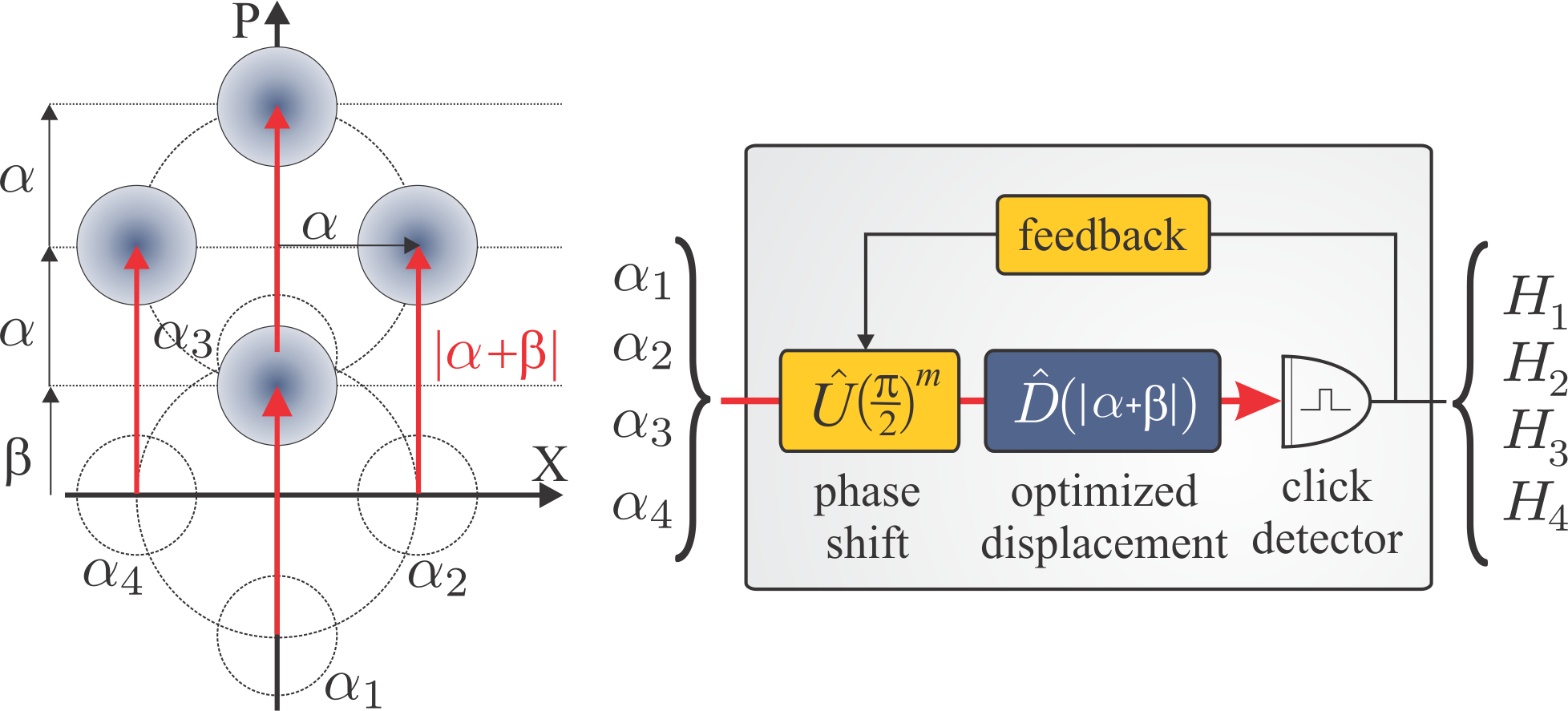}  %
\caption{(color online) (left) Illustration of the QPSK alphabet in phase space after applying the optimized displacement. (right) Sketch of the discrimination scheme. The states from the QPSK alphabet are optimally displaced prior to an on-off detector. The record of detection events is processed and fed back to a phase shifter to change the probed state.}
\label{BondurantOD}
\end{figure}

We will discuss two different types of optimized displacement receivers, differing in the order in which the signal states are probed. 
On the one hand we will investigate a direct generalization of the Bondurant receiver, where the probing of the states is carried out in a cyclic order $\alpha_1 \rightarrow\alpha_2 \rightarrow \alpha_3 \rightarrow \mathrm{\dots}\,$ . On the other hand, we will consider a more advanced probing strategy based on Bayesian inference, where the displacement after each detection event is adjusted to probe the state with the momentarily highest \textit{a posteriori} probability.

In both realizations, the sequential probing of the signal states can be described by the operator sequence $\hat{U}\left(\frac{2\pi}{M}\right)^m \, \hat{D}\left(|\alpha+\beta|\right)$, where $\hat{D}\left(\cdot\right)$ is the displacement operator,
$M$ denotes the number of signal states in the alphabet, e.g $M=4$ in the case of QPSK, 
and $\hat{U}\left(2\pi/M \right)^m = \exp\left(- i\,m\,2\pi/M\,\hat{n} \right)$ corresponds to a m-fold $2\pi/M$ phase shift.
The sequential probing of the signal states can hence be interpreted as a sequence of $2\pi/M$ phase shifts in the alphabet's original phase space configuration followed by the optimized displacement.
The two probing strategies merely differ in the values of the phase shift parameter $m$. In the cyclic case $m=1$ is a constant, whereas in the scheme with Bayesian probing $m \in \{1,2,\mathrm{\dots},M-1\}$ depends on the exact number and detection times of all previous detection events as will be discussed in more detail later. 
A phase space representation of an optimally displaced QPSK alphabet as well as a schematic of the receiver are depicted in Fig.\ref{BondurantOD}. 

In contrast to schemes that perform an exact nulling of the signals, the total number of detection events in the optimized displacement scheme is in general unbounded. 
In case of the cyclic probing strategy $\alpha_1 \rightarrow\alpha_2 \rightarrow \alpha_3 \rightarrow \mathrm{\dots}$, the initial displacement configuration is recovered after each multiple of $M$ detection events. 
The final hypothesis for the input state $\ket{\alpha_k}$ is hence associated with the detection of $\left(k-1\right)+M\,j$ photons, where $j \in N_0$.
The successful detection of the state $\ket{\alpha_2}$ in the QPSK protocol, for instance, is associated with the detection of $\left\{1,5,9,\mathrm{\dots}\right\}$ photons.
The average error probability follows as
\begin{equation}
P_{err}(|\alpha|) = 1 - \frac{1}{M} \sum_{k=1}^M  \sum_{j=0}^\infty P_{(k-1)+M\,j} \left(\alpha_k,\beta \right), 
\end{equation}
where $P_{(k-1)+M\,j} \left(\alpha_k, \beta \right)$ denotes the probability for the optimized displacement receiver to detect $(k-1)+ M\,j$ photons given the signal state $\ket{\alpha_k}$ and the displacement amplitude $\beta$.
In the Bayesian probing strategy the average error probability can only be evaluated numerically as the probing sequence is recursively depending on the series of detection times. 

In order to derive the error probabilities of the optimized displacement receivers explicitly, let us first discuss the photon detection statistics of coherent states in general terms. 
For an arbitrary coherent state $\ket{\alpha}$ with mean photon number $n = |\alpha|^2$, the time-independent photon statistics is described by a Poisson distribution.
The probability to observe exactly $m$ photons is given by 

\begin{equation}
P_{m}(n) = \frac{n^m}{m!}\,e^{-n}   
\label{Poisson}
\end{equation}

For a coherent state with a rectangular pulse shape (at unit temporal extent T=1), the relation between the signal's mean photon number and the measurement time $0\leq t \leq 1$ is linear.
The time-dependent statistics are therefore readily obtained by introducing the temporally resolved mean photon number $n(t)= n\cdot t$.
The probability of observing exactly one photon up to a measurement time $t$, for instance, is given by $P_1(n,\,t)=(n\,t)\,e^{-n\,t}$, where the photon could have been detected at any time $\tau$ within the interval $0<\tau<t$. 
The probability density for the photon to be detected exactly at time $t$ is given by the temporal density of not observing the vacuum state at time $t$
\begin{eqnarray}
p_{1}(t\,|\,n)&=& \frac{\partial}{\partial t }\left(1 - e^{-n\,t}\right) \nonumber \\ 	
						&=&  n\,e^{-n\,t}. 				
\label{clickdensity}
\end{eqnarray}

The probability to observe exactly one detection event in the complete measurement of the state ($t=1$) follows by integration over all possible detection times $0 \leq t_1 \leq 1$.
\begin{eqnarray}
P_{1}(n) &=& \int_0^1 n\,e^{-n\,t_1} \cdot e^{-n(1-t_1)}\, \mathrm{d}t_1 \nonumber \\
         &=&  n\,e^{-n},  
\label{oneclick}
\end{eqnarray}

where the second term $e^{-n(1-t_1)}$ corresponds to the probability of not observing any further photons after the first detection at $t_1$, i.e. the vacuum probability for the remainder of the state. 

Eq.(\ref{oneclick}) can readily be generalized to retrieve the probability to detect $m$ photons in the course of the complete measurement.
The corresponding equation comprises $m$ nested integrals,

\begin{equation}
\begin{split}
P_{m}(n)& = \int_0^1 \int_{t_1}^1 \cdots \int_{t_m}^1 (n\,e^{-n\,t_1})\cdot(n\,e^{-n\,(t_2-t_1)})\cdots  \\
        &(n\,e^{-n\,(t_{m} - t_{m-1} )})\cdot(e^{-n(1-t_m)})\,\mathrm{d}t_m \cdots\mathrm{d}t_2\,\mathrm{d}t_1\,,
\end{split}
\label{mclicks}
\end{equation}
where the limits of integration are chosen such that the detection of the $k$-th photon can only occur at times after the previous $k$-1 photons have been detected.
Eq.(\ref{mclicks}) reproduces the complete Poissonian photon statistics of the coherent state as given by Eq.(\ref{Poisson}).

In the optimized displacement receivers the argument of the phase shifter, and hence the probed state, is changed after each photon detection, such that the resulting photon number statistics will differ from a Poissonian distribution.
Based on Eq.(\ref{mclicks}), however, the corresponding statistics can be derived straightforwardly by replacing the fixed mean photon number $n$ by $n_{[k]}$, which denotes the respective mean photon number of the state after $k$ photon detections. 
The specific values of $n_{[k]}$ depend on the implemented probing strategy. 
In case of the cyclic probing, the probed signal changes sequentially according to the modulus relation $n_{\left[k\right]} = n_{1+\left(k\,\mathrm{mod}\,M\right)}$.

In the receiver scheme with Bayesian probing, however, the order in which the signals are probed is defined recursively by the number of previously detected photons and their corresponding detection times, i.e. after each detection event the signal with the highest \textit{a posteriori} probability needs to be calculated. 
After the detection of the first photon at time $t_1$, the \textit{a posteriori} probability for the state $\ket{\alpha_k}$ with mean photon number $n_k$ is given by 
\begin{eqnarray}
P(\alpha_k\,|\,t_1) &=& \frac{p(\,t_1\, |\,  n_k)}{\sum_{j=1}^M p(\,t_1\, |\,  n_j)}    \\
										&=& \frac{ n_k\,e^{-n_k\, t_1} }{ \sum_{j=1}^M n_j\, e^{-n_j\, t_1} }, \nonumber
\end{eqnarray}
where $p(\,t_1\,|\,n_k)$ is the probability density of the state $\ket{\alpha_k}$ to excite a detection event exactly at time $t_1$ as given by Eq.(\ref{clickdensity}).

Consequently, the \textit{a posteriori} probability for the signal state $\ket{\alpha_k}$ after the observation of $m$ photons at times $\left\{t_1, t_2,..., t_m\right\}$ is recursively given by 
\begin{equation}
P(\alpha_k\,|\, {t_1,...,t_m}) = \frac{P(\alpha_k\, |\,  {t_1,...,t_{m-1}}) \cdot p(\Delta t_m\, |\,  n_k)}{\sum_{j=1}^M P(\alpha_j\,|\,{t_1,...,t_{m-1}}) \cdot p(\Delta t_m\, |\,n_j)} ,  \nonumber
\label{a-posteriori}
\end{equation}
where $\Delta t_m = t_m-t_{m-1}$ and $t_0 = 0$. 

To achieve the minimal error probabilities, the displacement $|\beta|^2$ has to be optimized according to the given mean photon number of the original signal states $|\alpha|^2$. 
The optimized displacement parameter $\beta_{opt}$ is derived as the root of the error probability's derivative with respect to the displacement amplitude $\beta$.

\begin{equation}
\frac{\partial}{\partial \beta} P_{err}(\alpha, \beta)|_{\beta_{opt}} = 0
\end{equation}

\begin{figure}[htb]
\includegraphics[width=8cm]{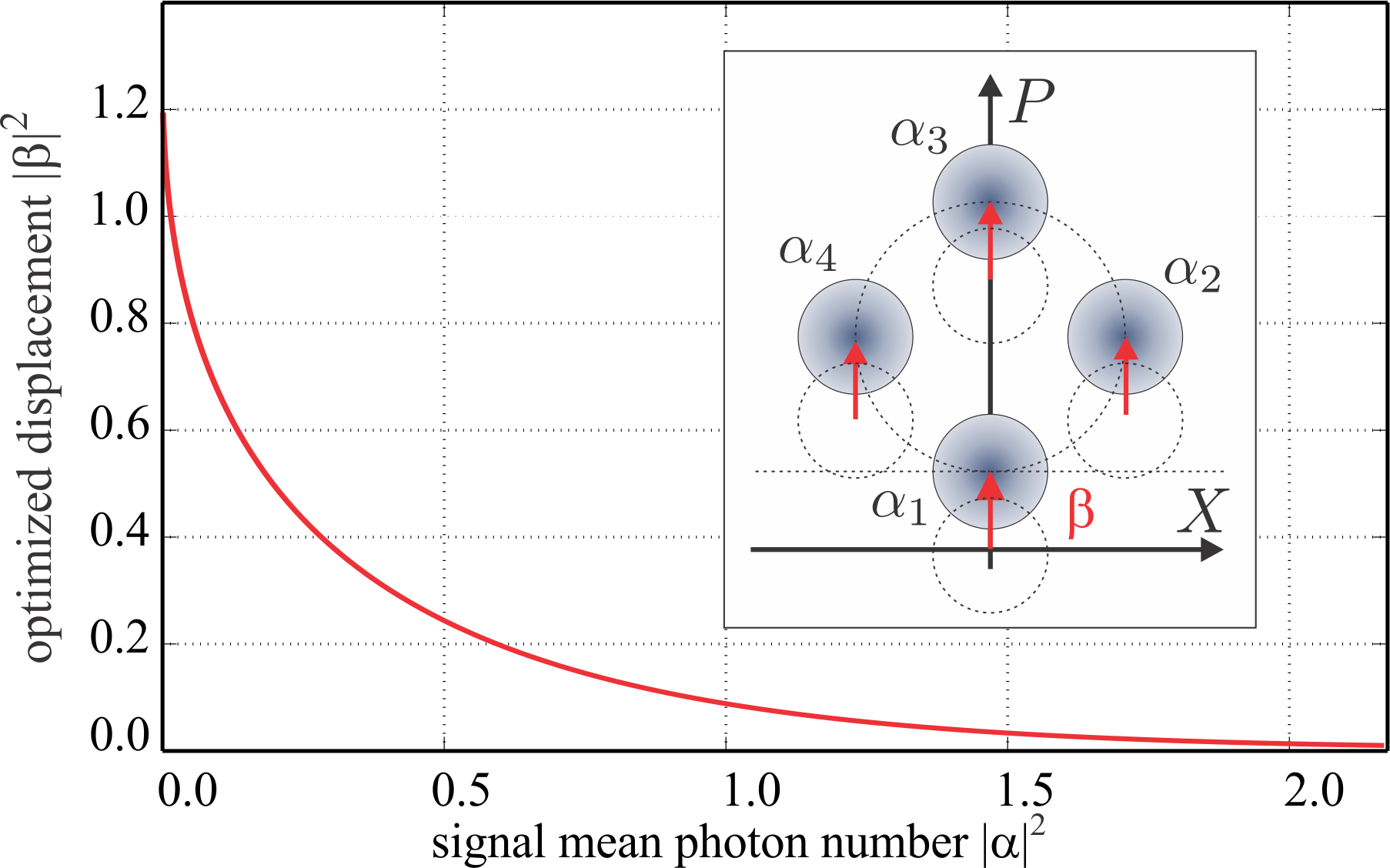}  
\caption{(color online) Displacement parameters to achieve the minimal error probabilities in the QPSK optimized displacement receiver with cyclic probing. For large signal amplitudes, the optimized displacement tends to zero such that the scheme approaches the Bondurant receiver.}
\label{optBETA}
\end{figure}

The optimized displacement parameters for the QPSK alphabet are shown in Fig.\ref{optBETA}. With decreasing signal power the value of the optimized displacement is continuously increasing and approaches a displacement amplitude corresponding to mean photon number as high as $1.2$ photons in the regime of extremely weak signals $(|\alpha|\approx 0)$. 
The optimized displacement actuates the probing sequence such that the bias on the initially probed state is extenuated. 
With increasing signal power, the photon detection probability for input states which are momentarily not displaced to the probing configuration, i.e. closest to the vacuum state, approaches unity. 
Consequently, the optimized displacement parameter tends to zero and the scheme approaches the configuration of the Bondurant receiver, i.e. the configuration corresponding to the exact nulling of the signal states. 

It is worthwhile mentioning that the clear structure of Eq.(\ref{mclicks}) is based on the Poissonian photon statistics inherent to any (pure) coherent state. 
Any detection event is independent of the record of prior detections. 
If any excess noise was present, the associated super-Poissonian statistics, i.e. the photon bunching effect, would render the individual detection probabilities mutually dependent and the problem would be considerably harder to grasp analytically. 
On first sight, it may seem as if the adaptive displacement essentially does add some kind of noise to the signal states. 
This is true, however, only as long as one thinks of the signal pulse as an indivisible unit. 
It is an important characteristic of coherent states that their quantum properties, i.e. their coherence, is preserved upon splitting on a beam splitter. 
The receiver actually performs a temporal splitting of the coherent input state conditioned on the detection times of the individual photons. 
Each interval between two detection events can be treated separately, such that the effectively detected signal can well be described by a product state of the form 
\begin{equation}
\rho\left(\{t_m\}\right) = \bigotimes_{k=0}^m \Delta t_{k} \ket{\sqrt{\Delta t_{k}}\,\alpha_{[k]}} \bra{\sqrt{\Delta t_{k}}\,\alpha_{[k]}},
\end{equation}
where $\alpha_{[k]}$ denotes the complex signal amplitude in each interval and $\Delta t_{k}$ denotes the temporal length of the interval.

\begin{figure}[htb]
\includegraphics[width=8cm]{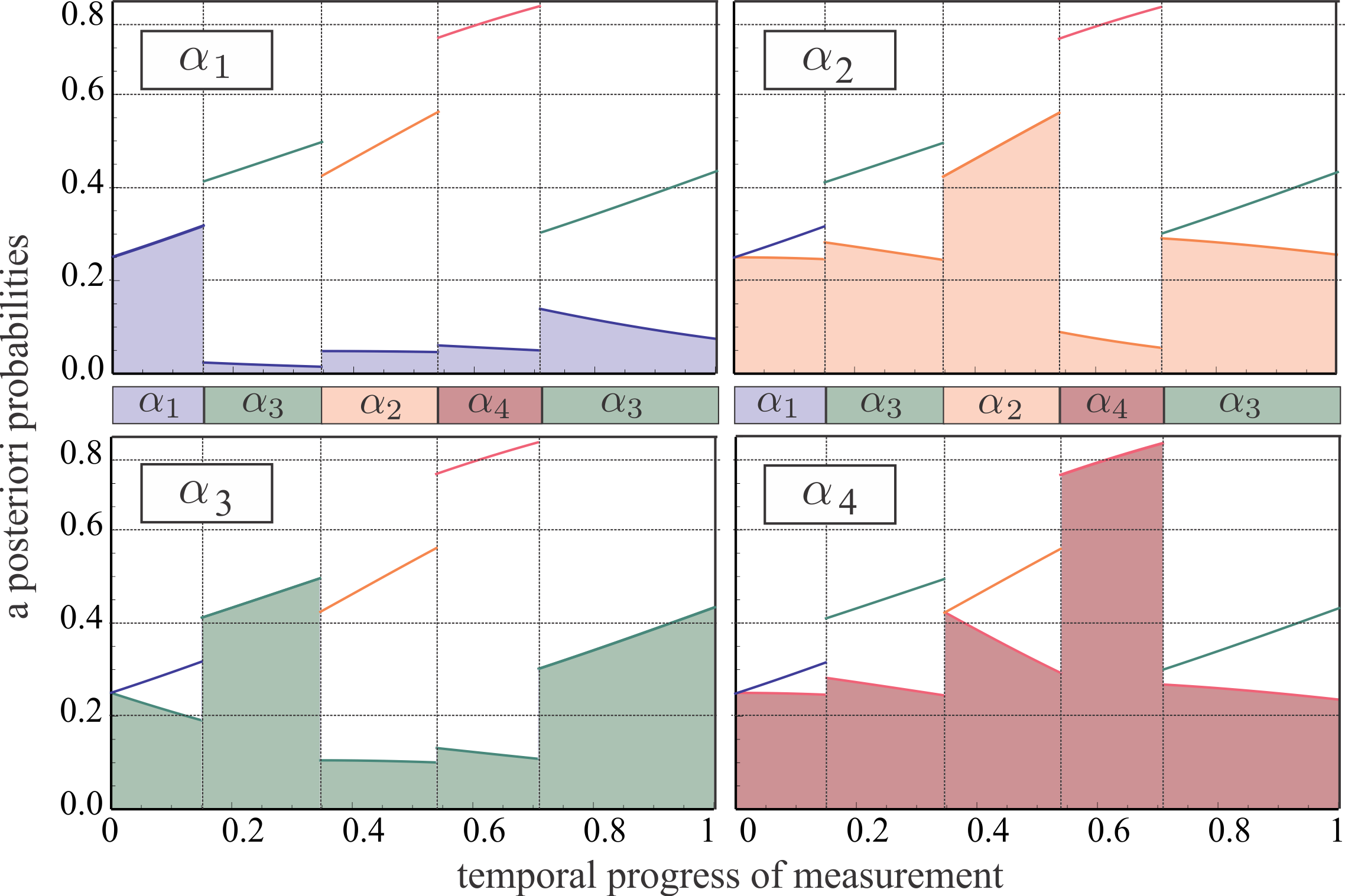}  %
\caption{(color online) Charts illustrating the changes in the \textit{a posteriori} probabilities for the successful detection of the signal state $\ket{\alpha_3}$ with mean photon number $|\alpha|^2=0.5$ and displacement $|\beta|^2=0.23$. In each graph, the states' instantaneous \textit{a posteriori} probability (filled curve) is shown together with the momentary maximal value (solid line). The change in the hypotheses of the receiver during the measurement process is underpinned by the bar between the two rows of plots, which displays the most likely state in each interval.}
\label{aposteriori}
\end{figure}

An example for the progression of the individual \textit{a posteriori} probabilities for the measurement of a QPSK signal with Bayesian probing is shown in Fig.\ref{aposteriori}. 
The graphs depict the successful identification of the signal state $\ket{\alpha_3}$ after the detection of four photons during the measurement of the complete signal pulse. 
The mean photon numbers of the displaced QPSK states are 
\begin{eqnarray}
n_1 &=& |\beta|^2,\,\,\,\,\,\,\,\,\,\,\,\,\,\, \,\,\,n_2 = |\alpha+\beta|^2+|\alpha|^2, \nonumber \\
n_3 &=& |2\alpha+\beta|^2, \,\,\,n_4 = n_2.  
\label{meanphotonnumbers}
\end{eqnarray}
The signal intensity for the exemplary measurement in Fig.\ref{aposteriori} is $|\alpha|^2=0.5$ and the value of the optimized displacement is $|\beta|^2=0.23$. 
Starting from symmetric \textit{prior} probabilities at $t=0$, the \textit{a posteriori} probability for the initially probed state $\ket{\alpha_1}$ is increasing until a first detection event is registered. 
In this particular example a photon detection at times $t_1<0.38$ maximizes the probability for $\ket{\alpha_3}$, whereas any later detection would render $\ket{\alpha_{2,4}}$ the most likely states after the first detection event. 
Given the detection time $t_1 = 0.15$, the updated \textit{a posteriori} probability thereafter favors the state $\ket{\alpha_3}$, which therefore is probed next. 
After three more detections, the totaling \textit{a posteriori} probability yields the final hypothesis for $\ket{\alpha_3}$. 
The exact values of the detection times and the corresponding \textit{a posteriori} probabilities of this example measurement are summarized in Tab.\ref{tabapost}.

\begin{table}[ht]
\caption{\textit{a posteriori} probabilities}  
\centering          
\begin{tabular}{c c c c c c c}    
\hline\hline 
Signal & $t_1$ = 0.15 & $t_2$ = 0.35 & $t_3$ = 0.54 & $t_4$ = 0.71 & $t=1$  \\[0.5ex]  
\hline   
$\alpha_1$ &  0.024 & 0.049 & 0.061 & 0.139 & 0.075 \\     
$\alpha_2$ &  0.277 & 0.423 & 0.090 & 0.291 & 0.256 \\
$\alpha_3$ &  0.403 & 0.105 & 0.131 & 0.302 & 0.434 \\
$\alpha_4$ &  0.277 & 0.423 & 0.718 & 0.268 & 0.236 \\[0.5ex]  
\hline         
\label{tabapost}
\end{tabular}
\end{table}

\begin{figure}[htb]
\includegraphics[width=8cm]{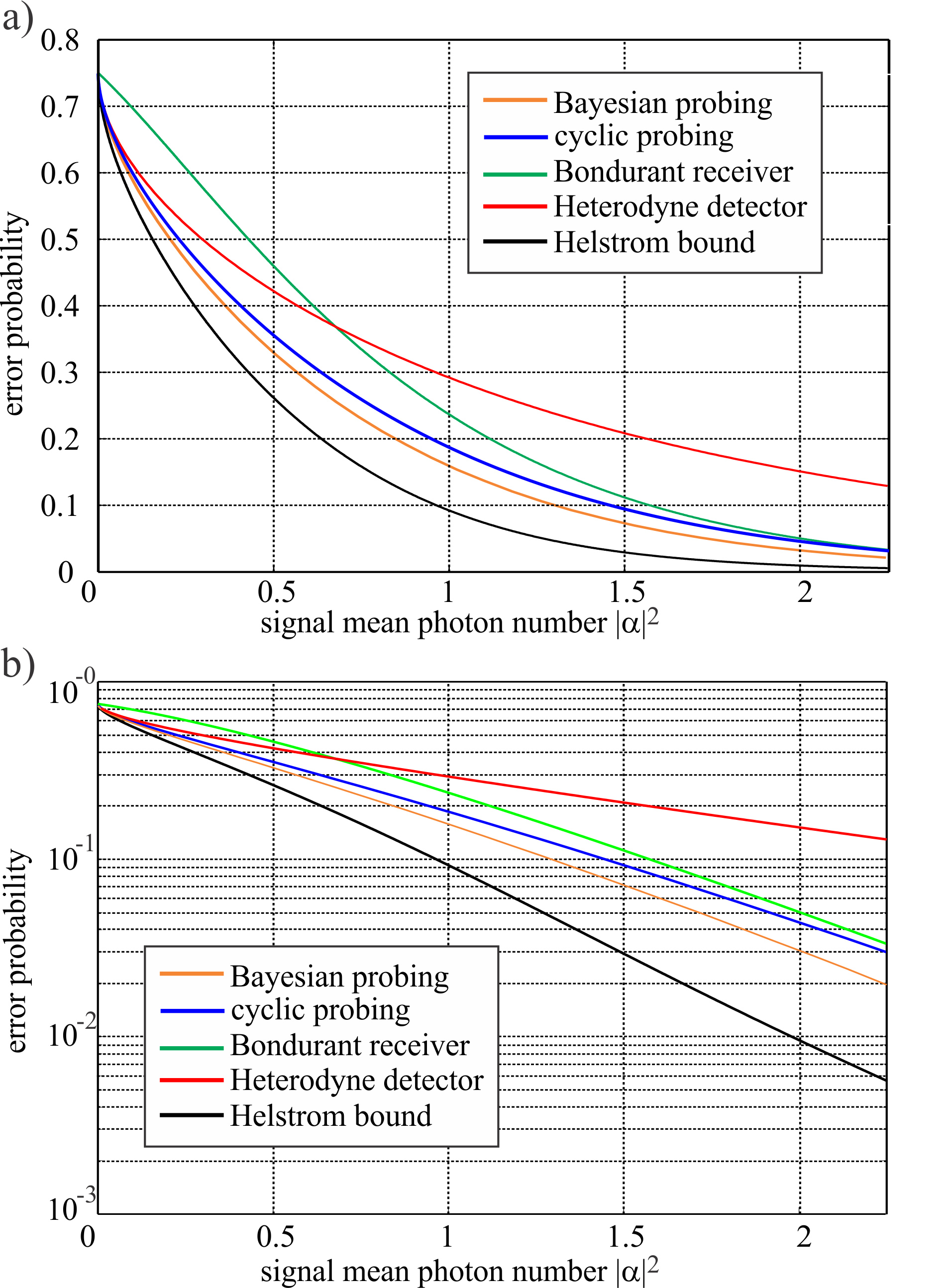}  %
\caption{(color online) Comparison of the error probabilities of the QPSK optimized displacement receivers with cyclic and Bayesian probing. Super-SQL performance 
is achieved for any signal amplitude: a)linear plot, b)logarithmic plot}
\label{curves}
\end{figure}

The error probabilities of the QPSK optimized displacement receivers have been evaluated by numerical Monte Carlo simulations and are shown and compared to the Bondurant receiver in Fig.\ref{curves}. Additionally, the Helstrom bound as well as the standard quantum limit, as achieved by an ideal heterodyne detector, are shown to serve as references. 
The error probabilities of the optimized displacement receivers are clearly below the SQL for any signal power. 
The benefit of the optimization of the displacement, i.e. the advantage over the Bondurant scheme, is most pronounced in the regime of weak signals.
Furthermore, the comparison between the cyclic and the Bayesian probing strategy points out that the super-SQL performance in the weak signal regime is mainly a consequence of the displacement optimization. The advantage of the Bayesian approach becomes pronounced only with increasing signal power, which is also underlined by the logarithmic representation. 
With increasing signal power, the curves of the optimized displacement scheme with cyclic probing and the curve of the Bondurant receiver coincide asymptotically as the optimized displacement parameter tends to zero. In contrast, the performance of the Bayesian probing approach is superior to the Bondurant receiver  at any signal power.

\begin{figure}[htb]
\includegraphics[width=8cm]{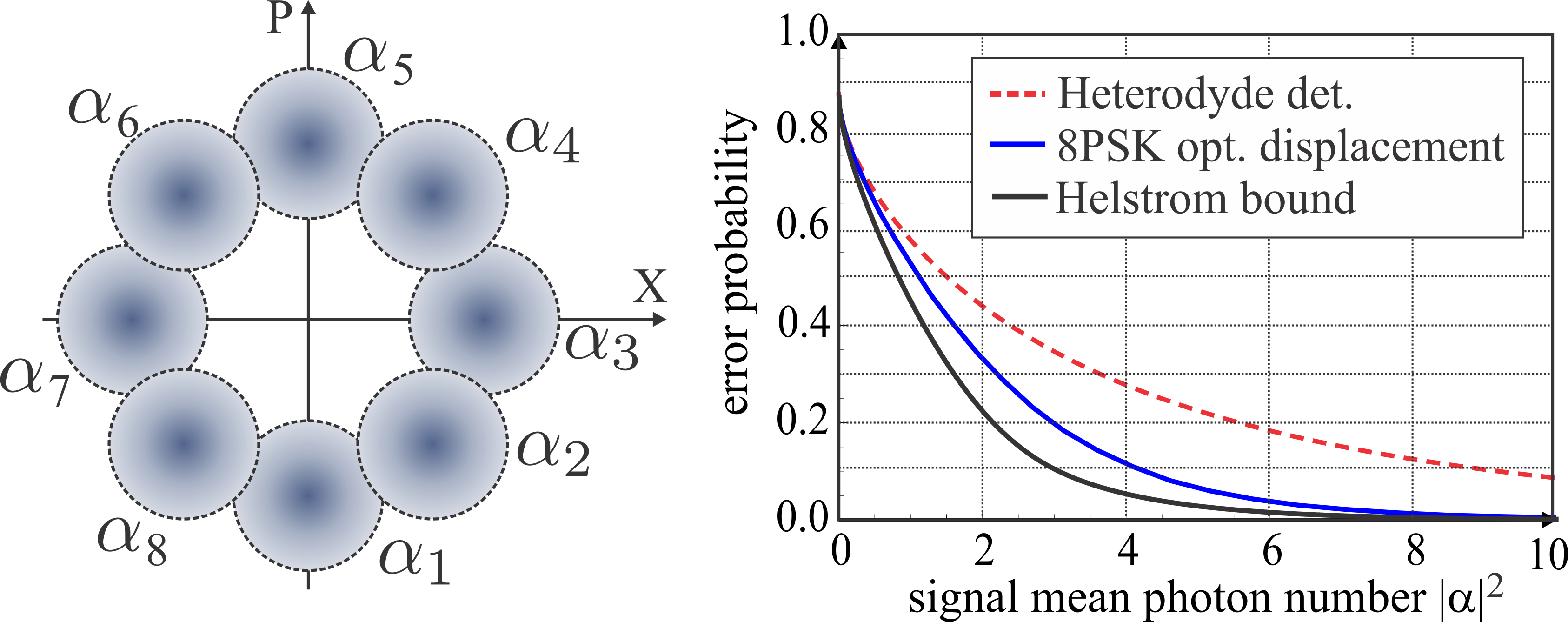}  %
\caption{(color online) (left) Illustration of the 8PSK alphabet in phase space. (right) Error probabilities of the 8PSK optimized displacement receiver (with Bayesian probing) compared to the standard quantum limit (heterodyne detection) and the Helstrom bound.
Just like for the QPSK alphabet, the standard quantum limit is outperformed for any signal power. }
\label{8PSK}
\end{figure}

We analyzed the performance of the optimized displacement receivers for phase shift keyed alphabets comprising up to eight signal states (8PSK). 
Qualitatively, all evaluated error probability curves resemble each other, such that the standard quantum limit is outperformed for any alphabet and for any signal power. 
We strongly conjecture that this capacity is also existent for any alphabet comprising more than eight signal states.
The resulting error probability for the 8PSK alphabet is depicted in Fig.\ref{8PSK}. 

\section{III. Robustness against experimental imperfection}
Up to this point all components of the channel and the receiver scheme were assumed to be ideal. 
In any practical realization, however, various imperfections can constrain the performance of the receiver and may eventually result in error probabilities above those of the standard quantum limit. 
In this section we will investigate the robustness of the QPSK optimized displacement receivers against four typical imperfections: finite quantum efficiency, excess noise, detector dead time and dark counts. 
In the corresponding sections, each imperfection is investigated separately via numerical Monte Carlo simulations, which allows to compare the impacts of the individual imperfections. 
Combinations of different imperfections may result in a more complex behavior which is subject to further investigations.

\subsection*{1.) Finite quantum efficiency}
Nowadays, the quantum efficiency of PIN photo diodes can be very close to unity $\eta\approx1$ such that the effectively detected signal can be very close to the actual input signal $\left|\tilde{\alpha}\right|^2 = \eta \left|\alpha\right|^2 \approx \left|\alpha\right|^2$. 
This is one of the reasons why homodyne and heterodyne detection, which both detect the signal via pairs of PIN photo diodes, are indeed adequate receiver schemes for many conventional applications. 
Still, a high detection efficiency in these receivers also requires a high interference contrast, i.e. interferometric visibility $VIS = (I_{max}-I_{min})/(I_{max} + I_{min})$, between the signal and the local oscillator, where $I_{max}$ ($I_{min}$) denotes the maximal (minimal) intensity in the corresponding interference oscillations. 
Finite visibility reduces the overall efficiency as $\eta_{HetD} = \eta_{diodes}\cdot VIS^2$ \cite{Silberhorn03}. 
In the following discussion, however, we shall assume that an ideal implementation of a heterodyne detector is indeed feasible. 
In contrast to PIN photo diodes, the typical quantum efficiency of single photon detectors, e.g. implemented by avalanche photo diodes, is by now still significantly smaller than unity $\eta<1$. 

In contrast to the heterodyne detector, the finite quantum efficiency in the optimized displacement receivers does not result in a direct attenuation of the signal as $\eta \left|\alpha_i\right|^2$, but rather on the displaced signals $\eta\, n\left(\alpha_k, \beta\right)$ as given by Eq.(\ref{meanphotonnumbers}). 
As a consequence, the ratio between the signal power and the displacement strength of the effectively detected signals is typically not optimized. 
However, this can straightforwardly be compensated by properly adjusting the displacement strength according to the specific value of the detectors quantum efficiency. 
In this case the consequence of finite quantum efficiency can be interpreted as an effective rescaling of the mean photon number as $|\alpha|^2 \rightarrow \eta |\alpha|^2$. 

\begin{figure}[htb]
\includegraphics[width=8cm]{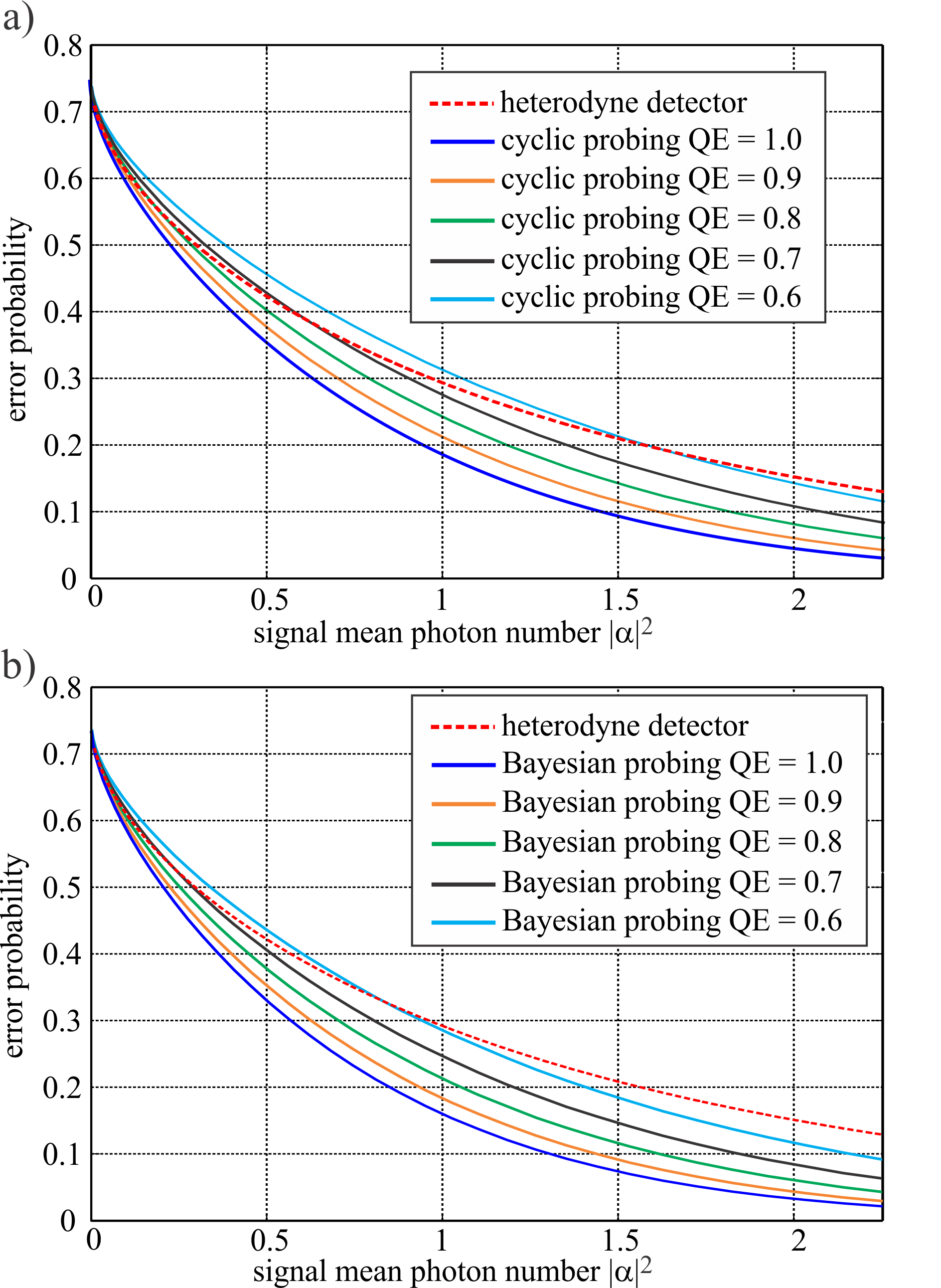}  %
\caption{(color online) Comparison of the QPSK optimized displacement receivers' performance for various quantum efficiency parameters: a)cyclic probing b)Bayesian probing. }
\label{QE}
\end{figure}

In Fig.\ref{QE} we compare the optimized error probabilities for the displacement receivers with different quantum efficiencies to those of the heterodyne detector with unit quantum efficiency and perfect visibility. The receiver with cyclic probing achieves super-SQL performance for most signal powers even with a quantum efficiency as low as $QE \approx 70\%$. 
However, even with lower efficiencies super-SQL performance can be achieved above a certain threshold signal power. 
The Bayesian probing strategy proves to be even more robust and super-SQL performance at a signal power of $|\alpha|^2 \approx 1$ can be achieved even for a quantum efficiency as low as $60\%$.

\subsection*{2) Excess noise}
In the discussion of optical communication protocols the channel is oftentimes assumed to merely act as an attenuator. 
For optical fibers, the channel loss is typically considered to be $0.2\,$dB/km for signals at wavelengths around 1550nm (C band). 
However, this model does usually not provide a complete description of the channel. 
Several interaction processes such as Raman- and Brillouin scattering, reflections at surfaces, (non-linear) crosstalk at wavelength division multiplexing channels or amplifier noise, couple the optical signal to the environment resulting in an unavoidable noise penalty onto the amplitude and phase properties of the signals. 
For the special case of binary phase shift keying (BPSK) and pure phase diffusion a recent work showed that homodyne detection can be a near optimal strategy \cite{ParisPhase}.
In general, the footprint of the excess noise depends on the specific channel parameters, but can in most cases be approximated by a thermal distribution. 
In phase space, the thermal noise contribution can be visualized as the result of a convolution between the original signal states and a symmetric Gaussian function. 
Using the convention $\hat{X} = \frac{1}{2}(\hat{a}+\hat{a}^{\dagger})$ for the definition of the field quadratures, the variance of this Gaussian function in terms of the mean number of thermal noise photons is given as $\sigma^2 = N_{th}/2$.

\begin{figure}[htb]
\includegraphics[width=8cm]{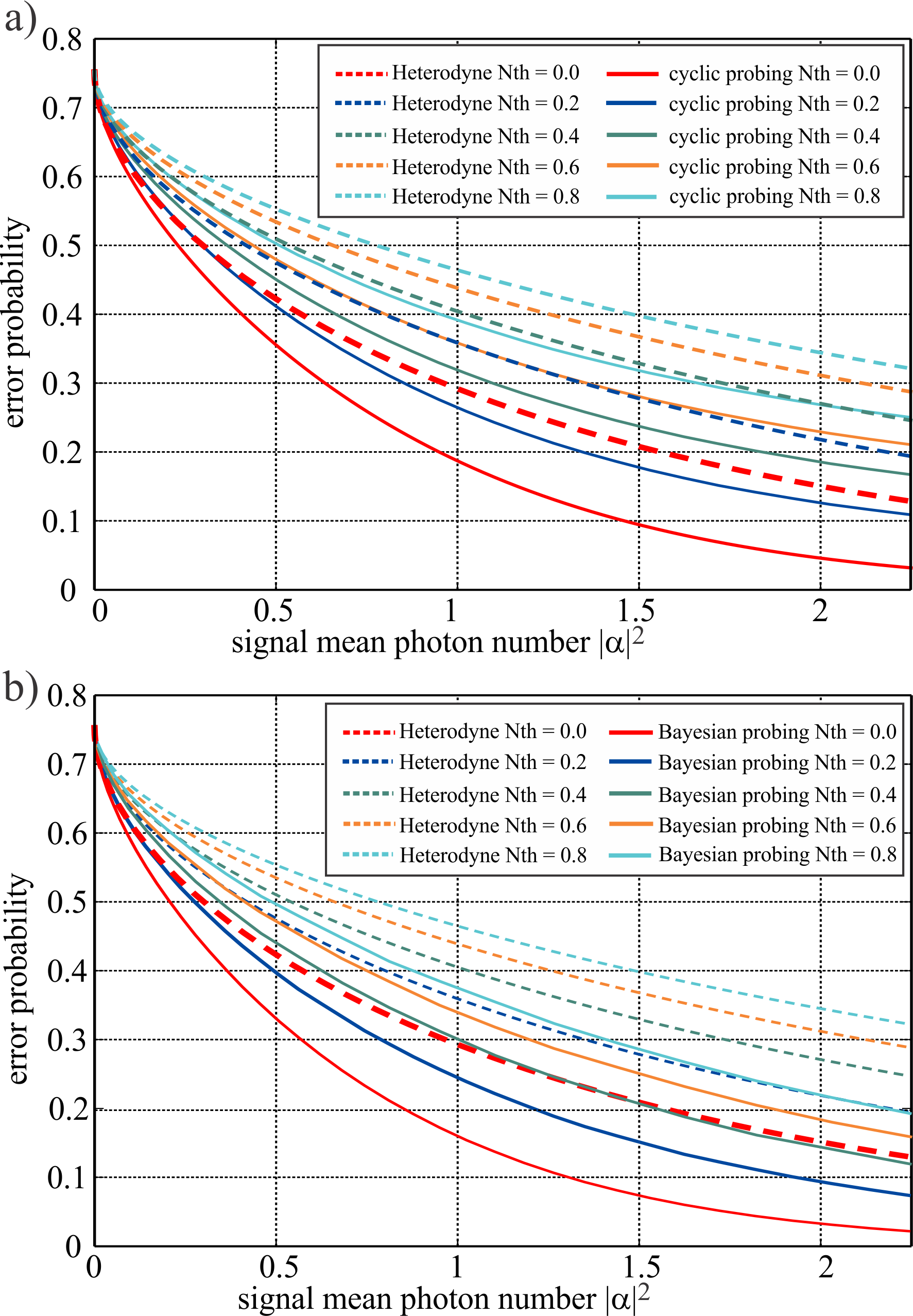}  %
\caption{(color online) Comparison of the error probabilities of the QPSK optimized displacement receivers and the heterodyne detector for the discrimination of thermal noise afflicted signals. Solid lines correspond to the optimized displacement receivers. Dashed lines depict the resulting error probability for the heterodyne receiver. Lines of equal color correspond to the same excess noise level: a)cyclic probing b)Bayesian probing. }
\label{ExcessNoise}
\end{figure}

Excess noise results in an increased error probability both for the optimized displacement receiver and for the heterodyne detector. 
The effects that constrain the performance in these two different receiver architectures are, however, of a different nature. 
In the heterodyne receiver the additional errors stem from the broadened support of the signal states on the field quadratures onto which the states are projected.  
In the optimized displacement receiver, the error probability is increased by additional noise-induced photon detections that result in an untimely update of the displacement phase and hence disturb the probing procedure.
In the Monte Carlo simulation, we benchmark the receivers' performance against excess noise with an average thermal photon number of up to 0.8 photons per signal state. 
The noise was simulated by applying random displacements with a Gaussian distribution to the signal states. 
This corresponds to thermal noise with a bandwidth close to the signal repetition rate. 
The resulting error probabilities are depicted in Fig.\ref{ExcessNoise} and are compared to the performance of a heterodyne receiver at the same thermal noise level. 
Despite the presence of excess noise, both probing techniques prove to achieve error probabilities below those of the heterodyne detector for any signal amplitude, which is a remarkable results for a receiver based on the detection of discrete variables.  

\subsection*{3.) Detector dead time}
\begin{figure}[htb]
\includegraphics[width=8cm]{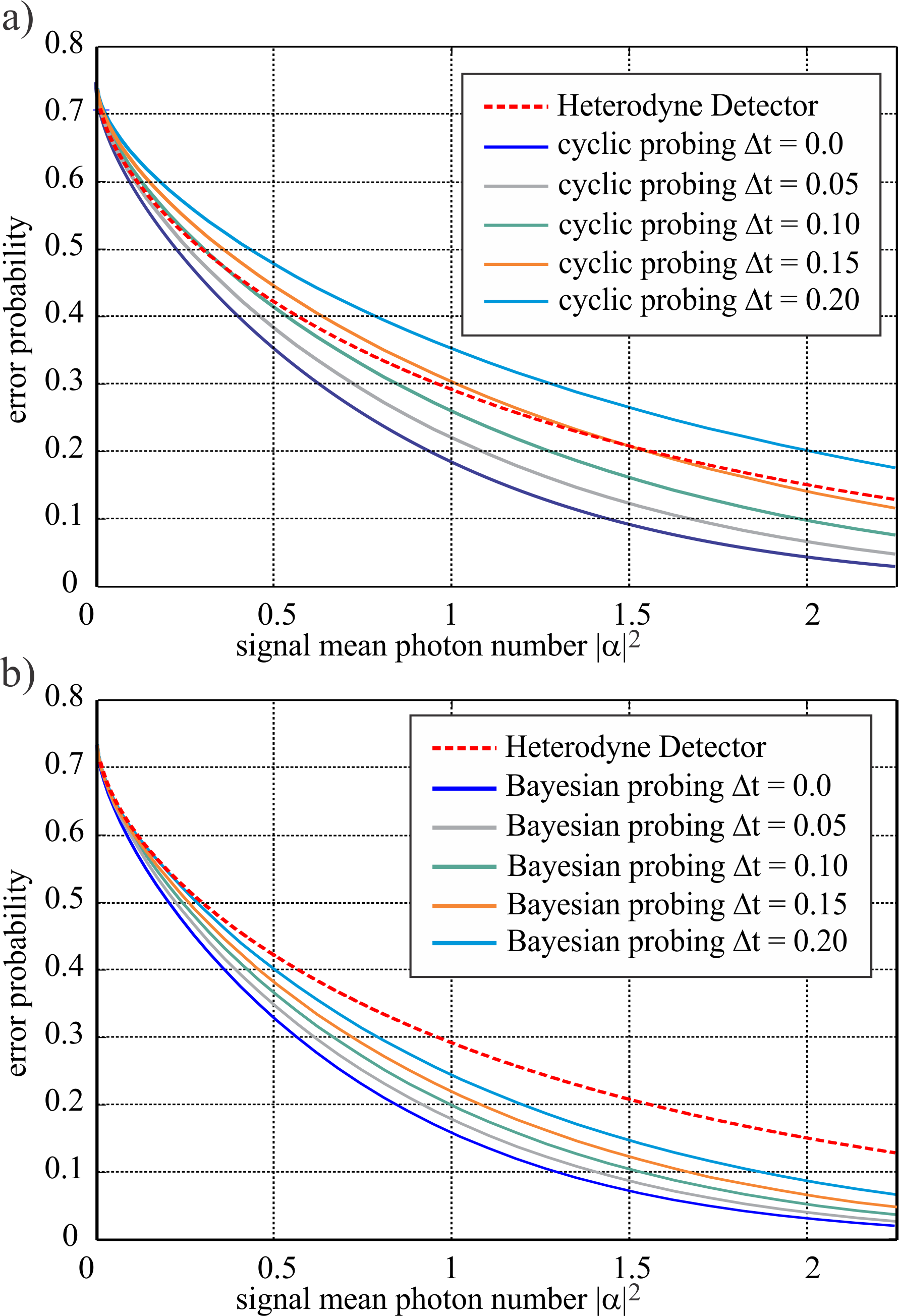}  %
\caption{(color online) Comparison of the error probabilities of the QPSK optimized displacement receivers and the ideal heterodyne detector for detector dead times masking up to 20$\%$ of the signal's temporal extent for each single photon detection: a)cyclic probing b)Bayesian probing.}
\label{Deadtime}
\end{figure}

In an avalanche photo diode a detection event is associated with a cascaded emission of electrons in the diode material. 
After each detection event, the resulting electron avalanche first needs to be quenched, before another photon can be detected. 
During this dead time the detector is blind to the signal. 
In this respect, dead time can be interpreted as a measurement induced on-off modulation of the detector. 
As the number of detection events in the optimized displacement receiver is in general unbounded, the detrimental effect of the dead time might have a significant impact onto the receivers' performance. 
In Fig.\ref{Deadtime} the error probability of the receivers are shown for different dead time parameters $\Delta t$, which indicate the percentage of the states' temporal extent blinded by a single photon detection. The investigated parameters range from dead time free detection ($\Delta t = 0.0$) up to the case where each single detection event masks 20$\%$ of the signal state ($\Delta t = 0.20$). 
The receiver with cyclic probing can tolerate dead times as high as $\Delta t \approx 0.10$ without loosing the super-SQL performance for most signal amplitudes. 
Yet, the receiver with Bayesian probing proves to be even more robust also with respect to the dead time and maintains the super-SQL performance even for a dead time parameter as high as $\Delta t = 0.20$. 
This characteristic reflects the fact that on average, the Bayesian probing procedure requires considerably less photon detections before the actual input state is probed.
Consequently, the integrated blind time of the detector is reduced and the receiver can tolerate a larger dead time per photon detection.

\subsection*{4.) Dark counts}
In conventional detectors with single photon sensitivity, e.g. avalanche photo diodes, the detector output is disturbed by the occurrence of erroneous, thermally excited events - dark counts. 
In the case of quantum receivers relying on a perfect nulling of the probed states, such events obviously have a strong impact on the final hypothesis and hence result in a significantly increased error probability. 
In the optimized displacement receiver, the detrimental effect of dark counts is extenuated. 
The dark count statistics can typically by described by a Poissonian distribution. 
In this case the dark count rate $n_{dc}$ will simply add to the photon count rate of the displaced optical signal $n_i\left(\alpha_i, \beta \right)$ and result in an overall increased mean photon number $\tilde{n}_i = n_i + n_{dc}$. 
In situations in which the actual input state is not yet probed, the occurrence of a dark count actually assists the state discrimination procedure in the sense that it triggers the switching of the probed state. 
In this sense the effect of the dark counts is similar to that of the optimized displacement itself.
However, in contrast to the dark counts, the displacement is an interference between the complex amplitude of the displacement field $\beta$ and the signal amplitude $\alpha_i$, such that the mean photon number of the displaced signals scales as $|\alpha+\beta|^2$, which provides the superior performance. 
Particularly, the mutual distance of the signal states in phase space is invariant under the displacement operation, i.e. the states are still arranged on a circle with radius $|\alpha|$. 
In contrast, the dark counts lead to a deformation of this circle such that in total, the mutual overlap of the signal states is increased. 

\begin{figure}[htb]
\includegraphics[width=8cm]{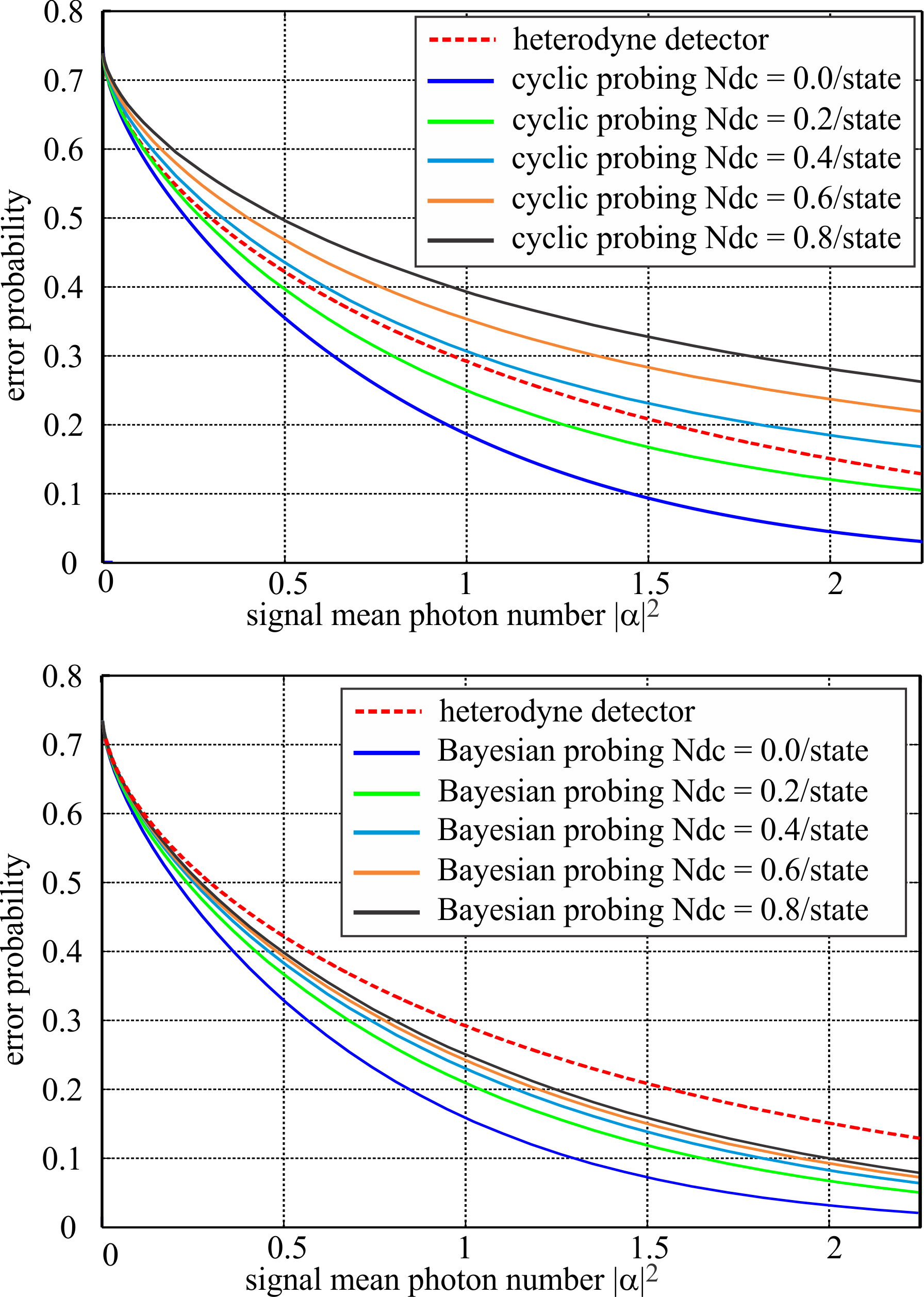}  %
\caption{(color online) Error probabilities for the QPSK optimized displacement receivers with dark count rates up to 0.8 dark counts per state on average: a)cyclic probing b)Bayesian probing.}
\label{darkcounts}
\end{figure}

In Fig.\ref{darkcounts}, we compare the error probability curves of the optimized displacement receivers for different dark count rates ranging up to an average of $0.8$ dark count detections per signal state. 
The Bayesian probing approach proves to provide significantly higher robustness compared to the cyclic probing scheme. 
In the cyclic probing scheme, the occurrence of an erroneous detection event at a time when the actual input state is already probed results in the requirement to detect three more photons before the input state is again considered as the final hypothesis. 
In the Bayesian probing scheme this is in general not the case.
As the Bayesian probing scheme keeps record of the \textit{a posteriori} probabilities, a dark count is typically compensated in fewer than three subsequent photon detections. 
Depending on the record of prior detections, the receiver optimally assesses the occurrence of the dark count and may eventually return to probing the actual input state yet after a single detection event. 

\section{IV. Conclusion}
In this article, we proposed an optimized quantum receiver scheme for the discrimination of phase shift keyed coherent signals. 
We compared two different probing strategies, a simple cyclic switching of the states and a more involved Bayesian probing approach.
We argued that independent of the number of signal states in the PSK alphabet, the receiver provides error probabilities below the standard quantum limit for any signal power. 
Moreover, the analysis regarding the robustness of the scheme against technical imperfections demonstrates that our quantum receiver can outperform any classical receiver scheme even under realistic conditions. 
As low error probabilities are crucial both to approach the channel capacity in optical communications and to increase the secret key rate in quantum key distribution protocols, we believe that our receiver will play a significant role in future communication and quantum information technologies.

\subsection*{Acknowledgments}  
The authors thank A.~Leven, H.~B\"{u}low and L.~Schmalen for fruitful discussions. 
This work was supported by Alcatel-Lucent Bell Labs Germany.

\bibliographystyle{unsrt}

\begin{thebibliography}{37}
\expandafter\ifx\csname natexlab\endcsname\relax\def\natexlab#1{#1}\fi
\expandafter\ifx\csname bibnamefont\endcsname\relax
  \def\bibnamefont#1{#1}\fi
\expandafter\ifx\csname bibfnamefont\endcsname\relax
  \def\bibfnamefont#1{#1}\fi
\expandafter\ifx\csname citenamefont\endcsname\relax
  \def\citenamefont#1{#1}\fi
\expandafter\ifx\csname url\endcsname\relax
  \def\url#1{\texttt{#1}}\fi
\expandafter\ifx\csname urlprefix\endcsname\relax\def\urlprefix{URL }\fi
\providecommand{\bibinfo}[2]{#2}
\providecommand{\eprint}[2][]{\url{#2}}


\bibitem{Giovannetti04}
V.~Giovannetti, S.~Guha, S.~Lloyd, L.~Maccone, J.~H.~Shapiro, and H.~P.~Yuen, 
{\em Classical Capacity of the Lossy Bosonic Channel: The Exact Solution}
Phys. Rev. Lett. \textbf{92}, 027902 (2004)

\bibitem{Weedbrook12}
C.~Weedbrook, S.~Pirandola, R.~Garc\'{i}a-Patr\'on, N.~J.~Cerf, T.~C.~Ralph, J.~H.~Shapiro, and S.~Lloyd
{\em Gaussian Quantum Information}
Rev. Mod. Phys. \textbf{84}, 621--669 (2012)

\bibitem{Scarani09}
V.~Scarani, H.~Bechmann-Pasquinucci, N.~J.~Cerf, M.~Du\u{s}ek, N.~L\"{u}tkenhaus, and M.~Peev
{\em The security of practical quantum key distribution}
Rev. Mod. Phys. \textbf{81}, 1301--1350 (2009)

\bibitem{Bennett92}
C.~H.~Bennett,
{\em Quantum cryptography using any two nonorthogonal states}
Phys. Rev. Lett. \textbf{68}, 3121-3124 (1992)

\bibitem{Lorenz04}
 S.~Lorenz, N.~Korolkova and G~Leuchs,
{\em Continuous-variable quantum key distribution using polarization encoding and post selection}
 Appl. Phys. B \textbf{79} (3), 273-277 (2004)
 
\bibitem{Sych10}
D.~Sych and G.~Leuchs, 
{\em Coherent state quantum key distribution with multi letter phase-shift keying}
New. J. Phys. \textbf{12}, 053019 (2010)
 
\bibitem{Leverrier09}
A.~Leverrier, and P.~Grangier,
{\em Unconditional Security Proof of Long-Distance Continuous-Variable Quantum Key Distribution with Discrete Modulation}
Phys. Rev. Lett.,  \textbf{102}, 180504 (2009)
 
\bibitem{Helstrom67}
C.~W.~Helstrom,
{\em Detection theory and quantum mechanics}
Inform. Control \textbf{10} (1967)

\bibitem{Helstrom76}
C.~W.~Helstrom,
{\em Quantum Detection and Estimation Theory} 
(Academic, New York, 1976) Chap. VI p. 187.

\bibitem{Bergou04}
J.~A.~Bergou, U.~Herzog, and M.~Hillery,
{\em Quantum State Estimation}, 
(Lect. Notes Phys. vol 649) ed. M.~Paris and J.~\v{R}eha\v{c}ek (Berlin: Springer) chapter 11 pp 417-465 (2004)

\bibitem{Dolinar73}
S.~Dolinar,
Research Laboratory of Electronics, MIT, Quarterly Progress Report No. \textbf{111}, p. 115  (1973)

\bibitem{Kennedy73}
R.~S.~Kennedy,
Research Laboratory of Electronics, MIT, Quarterly Progress Report No. \textbf{108}, p. 219 (1973)

\bibitem{HirotaBan96}
M.~Sasaki, M.~Ban, and O.~Hirota, 
{\em Derivation and physical interpretation of the optimum detection operators for coherent-state signals}
Phys. Rev. A \textbf{54}, 1691 (1996)	

\bibitem{Nair12}
R.~Nair and B.~J.~Yen, 
{\em Symmetric M-ary phase discrimination using quantum-optical probe states}
Phys. Rev. A \textbf{86}, 0223068 (2012)		

\bibitem{Banaszek99}
K.~Banaszek, 
{\em Optimal Receiver for quantum cryptography with two coherent states}
Phys. Lett. A \textbf{253}, 12-15 (1999)		
	
\bibitem{Bondurant93}
R.~S.~Bondurant, 
{\em Near-quantum optimum receivers for the phase-quadrature coherent-state channel}
Opt. Lett. \textbf{18}, 22 1896-1898 (1993)	
	
\bibitem{Becerra11}
F.~E.~Becerra, J.~Fan, G.~Baumgartner, S.~V.~Polyakov, J.~Goldhar, J.~T.~Kosloski, and A.~Migdall, 
{\em M-ary-state phase-shift-keying discrimination below the homodyne limit}
Phys. Rev. A \textbf{84}, 062324 (2011)	
	
\bibitem{Becerra13NatPhot}
F.~E.~Becerra, J.~Fan, G.~Baumgartner, J.~Goldhar, J.~T.~Kosloski, and A.~Migdall,
{\em Experimental demonstration of a receiver beating the standard quantum limit for multiple nonorthogonal state discrimination}
Nature Phot. \textbf{7}, 147-152 (2013)		

\bibitem{Izumi12}
S.~Izumi, M.~Takeoka, M.~Fujiwara, N.~Dalla Pozza, A.~Assalini, K.~Ema, and M.~Sasaki, 
{\em Displacement receiver for phase-shift-keyed coherent states}
Phys. Rev. A \textbf{86}, 042328 (2012)		

\bibitem{Becerra13NatComm}
F.~E.~Becerra, J.~Fan, and A.~Migdall,
{\em Implementation of generalized quantum measurements for unambiguous discrimination of multiple non-orthogonal coherent states}
Nature Comm. \textbf{4}, 2028 (2013)	

\bibitem{Assalini11}
A.~Assalini, N.~Dalla~Pozza, and G.~Pierobon
{\em Revisiting the Dolinar receiver through multiple-copy state discrimination theory}
Phys. Rev. A \textbf{84}, 022342 (2011) 

\bibitem{Sych13}
D.~Sych and G.~Leuchs
{\em Optimal receiver for binary coherent signals},
arXiv:1404.5033

\bibitem{Nair14}
R.~Nair, S.~Guha, and S.\.Tan
{\em Realizable receivers for discriminating coherent and multicopy quantum states near the quantum limit}
Phys. Rev. A \textbf{89}, 032318 (2014) 

\bibitem{Izumi13}
S.~Izumi, M.~Takeoka, K.~Ema, and M.~Sasaki, 
{\em Quantum receivers with squeezing and photon-number resolving detectors for \textit{M}-ary coherent state discrimination}
Phys. Rev. A \textbf{87}, 042328 (2013)		

\bibitem{Mueller12}
C.~R.~M\"{u}ller, M.~A.~Usuga, C.~Wittmann, M.~Takeoka, Ch.~Marquardt, U.~L.~Andersen, and G.~Leuchs, 
{\em Quadrature phase shift keying coherent state discrimination via a hybrid receiver}
NJP \textbf{14}, 083009 (2012)	

\bibitem{Hirota96}
M.~Sasaki and O.~Hirota, 
{\em Optimum decision scheme with a unitary control process for binary quantum-state signals}
Phys. Rev. A \textbf{54}, 2728 (1996)	
	
\bibitem{Geremia04}
J.~M.~Geremia, 
{\em Distinguishing between optical coherent states with imperfect detection}
Phys. Rev. A \textbf{70}, 062303 (1996)

\bibitem{Takeoka05}
M.~Takeoka, M.~Sasaki, P.~van~Loock, and N.~L\"{u}tkenhaus 
{\em Implementation of projective measurements with linear optics and continuous photon counting}
Phys. Rev. A \textbf{71}, 022318 (2005)

\bibitem{Takeoka06}
M.~Takeoka, M.~Sasaki, and N.~L\"{u}tkenhaus 
{\em Binary Projective Measurement via Linear Optics and Photon Counting}
Phys. Rev. Lett. \textbf{97}, 040502 (2006)

\bibitem{Takeoka08}
M.~Takeoka and M.~Sasaki
{\em Discrimination of the binary coherent signal: Gaussian-operation limit and simple non-Gaussian near-optimal receivers}
Phys. Rev. A \textbf{78}, 022320 (2008)

\bibitem{Wittmann08}
C.~Wittmann, M.~Takeoka, K.~Cassemiro, M.~Sasaki, G.~Leuchs and  U.~L.~Andersen,
{\em Demonstration of Near-Optimal Discrimination of Optical Coherent States}
Phys. Rev. Lett., \textbf{101}, 210501 (2008)

\bibitem{Sasaki13}
M.~Fujiwara, S.~Izumi, M.~Takeoka, M.~Sasaki
{\em Performance gain of displacement receiver with optimized prior probability}
arXiv:1308.2036 

\bibitem{Geremia07}
R.~L.~Cook, P.~J.~Martin and J.~M.~Geremia,
{\em Optical coherent state discrimination using a closed-loop quantum measurement},
Nature \textbf{446}, 774-777	(2007)
 
\bibitem{Tsujino10} 
K.~Tsujino, D.~Fukuda, G.~Fujii, S.~Inoue, M.~Fujiwara, M.~Takeoka, and M.~Sasaki
{\em Sub-shot-noise-limit discrimination of on-off keyed coherent signals
via a quantum receiver with a superconducting transition edge sensor}, 
Opt. Express, \textbf{18}, 8107  2010)

\bibitem{Tsujino11}
K.~Tsujino, D.~Fukuda, G.~Fujii, S.~Inoue, M.~Fujiwara, M.~Takeoka, and M.~Sasaki, 
{\em Quantum Receiver beyond the Standard Quantum Limit of Coherent Optical Communication}
Phys. Rev. Lett. \textbf{106}, 250503 (2011)		
	
\bibitem{Silberhorn03}
C.~Silberhorn, 
{\em Intensiv verschr\"{a}nkte Zust\"{a}nde und Quantenkryptographie.}
Ph.D. thesis, University of Erlangen-Nuremberg, Germany (2003)			
	
\bibitem{ParisPhase}
S.~Olivares, S.~Cialdi, F.~Castelli, and M.~Paris
{\em Homodyne detection as a near-optimum receiver for phase-shift-keyed binary communication in the presence of phase diffusion}, 
Phys. Rev. A, \textbf{87}, 050303 (2013)
	
	
\end{thebibliography}

\end{document}